# A nearly perfect omnidirectional shear-horizontal (SH) wave transducer based on a thickness poled, thickness-shear ($d_{15}$) piezoelectric ring

Qiang Huan[1,2], Hongchen Miao[1,2], and Faxin Li[1,2,*]

[1] LTCS and Department of Mechanics and Engineering Science, College of Engineering, Peking University, Beijing, 100871, China

[2] Center for Applied Physics and Technology, Peking University, Beijing, 100871, China

**Abstract:** The fundamental shear horizontal ($SH_0$) wave in plates is of great importance in the field of nondestructive testing (NDT) and structural health monitoring (SHM) since it is the unique non-dispersive guided wave mode. For practical applications, a phased array system based on omnidirectional $SH_0$ wave transducers is most useful as it can cover a wide range of a plate. However, so far very few omnidirectional SH wave transducers have been developed. In this work, we proposed an omnidirectional SH piezoelectric transducer (OSH-PT) based on a thickness poled piezoelectric ring. The ring is equally divided into twelve sectors and the electric field is circumferentially applied, resulting in a new thickness-shear ($d_{15}$) mode. Finite element analysis shows that the proposed OSH-PT can excite single-mode $SH_0$ wave and receive the $SH_0$ wave only. Experiments were then conducted to examine the performance of the proposed OSH-PT. Results indicated that it can generate and receive single-mode $SH_0$ wave in a wide frequency range with nearly uniform sensitivities along all directions. Considering its quite simple configuration, compact size and low cost, the proposed OSH-PT is expected to greatly promote the applications of SH waves in the field of NDT and SHM.

---

[*] Author to whom all correspondence should be addressed. Email: lifaxin@pku.edu.cn

# 1. Introduction

Nondestructive testing (NDT) and structural health monitoring (SHM) have become more and more important in modern industries. The periodical inspection of large plate-like structures is a useful way to avoid catastrophic failure and economic losses. Unlike the conventional time-consuming ultrasonic testing method, ultrasonic guided wave has been proven to be an effective technique for large-structure inspection due to its capability of wide range and long distance interrogation[1-3]. For practical applications, a phased array system composed by omnidirectional guided wave transducers is of most helpful[4]. Because the wave energy can be concentrated at any direction, this phased array system may cover all surrounding areas up to several meters using omnidirectional B-scan with uniform modal selectivity[5]. As the key elements of the phased array system, the omnidirectional guided wave transducers have been paid much attention. In recent years, several types of omnidirectional Lamb wave transducers have been proposed based on piezoelectric effect[6, 7], magnetostrictive effect[8] and electromagnetic acoustic principle[9, 10].

In plate-like structures, there exist two types of guided waves including Lamb waves and shear horizontal (SH) waves. Currently, Lamb waves have been widely used in both NDT and SHM. However, it is well known that Lamb waves are inherently multi-mode and dispersive. Due to the dispersion, signals will be distorted with increase of propagation distance, resulting in complicated signal processing. Compared with Lamb waves, the fundamental shear horizontal ($SH_0$) wave is completely non-dispersive and less mode conversion occurs when it encounters defects or boundaries, which reduces the complexity of signal processing. Besides, when detecting structures surrounded by non-viscous liquid, theoretically $SH_0$ is of non-attenuation except for the geometric spreading [11]. Despite of the above advantages of $SH_0$ wave, very few omnidirectional $SH_0$ wave transducers have been developed so far. Based on magnetostrictive effect, Seung et.al developed an omnidirectional SH wave magnetostrictive patch transducer (MPT) [12] in which the dynamic magnetic field is achieved by alternating current in a coil. Recently, Seung et.al proposed an omnidirectional SH wave electromagnetic acoustic transducer (EMAT) based on Lorentz force [13]. Obviously, this type transducer is only suitable for metallic. It should be noted that for both MPT and EMAT, the energy conversion efficiency is very low and the requirement of magnets results in a large footprint, which is not suitable for SHM.

Compared with MPT and EMAT, piezoelectric wave transducers are more promising in NDT and especially in SHM for their compact size and high energy conversion efficiency. Using conventional thickness-poled piezoelectric transducers, Lamb waves can be excited and detected via $d_{33}$ mode or $d_{31}$ mode. It is Wilcox et.al that firstly adopted the thickness-shear ($d_{15}$) mode piezoelectric ceramic to excite and receive SH waves[14]. Later, Kamal and Giurgiutiu systematically investigated the performance of $d_{15}$ mode piezoelectrics in exciting and receiving SH waves and found that Lamb waves will always be excited simultaneously[15]. Recently, Zhou et.al excited and received $SH_0$ wave in a plate using face-shear ($d_{36}$) mode piezoelectric single crystals[16]. More recently, Miao et al realized face-shear ($d_{36}$) mode and apparent $d_{36}$ mode in PZT ceramics, and successfully excited and received SH0 wave in plates [17-20]. However, the d36 mode piezoelectrics still cannot excite single-mode SH wave because the d31 mode always coexists in the transducer. A breakthrough was achieved by Miao et al recently that they proposed the face shear d24 mode piezoelectric transducer and successfully excited and received single-mode $SH_0$ wave in plates[21]. However, all the piezoelectric SH wave transducers mentioned above could not be directly used in a phased array system because the $SH_0$ wave can only be generated at specific directions.

Recently, there are also some attempts to develop omnidirectional SH wave piezoelectric transducers. Borigo et.al[22] proposed the design of an omnidirectional SH wave transducer composed by two circumferentially poled PZT hollow cylinders. However, it is almost impossible to realize a uniformly circumferentially poled PZT cylinder in practice. Belanger et.al[23] developed an omnidirectional SH wave transducer by using six triangle PZT wafers to form a circular array. Although the simulation result of their design is acceptable, the testing result on their fabricated transducer is not successful. Miao et.al[24] fabricated an omnidirectional SH wave transducer consisting of twelve $d_{24}$ trapezoidal PZT elements, and excited and received single-mode $SH_0$ wave at all directions with the sensitivity variations of about 15%. It should be noted that actually the above designed omnidirectional SH wave transducers were all based on circumferentially poled or equivalent piezoelectric rings, which bring difficulties in the fabrication process since it is almost impossible to realize uniform circumferential poling in PZT ceramics.

In this work, we proposed an omnidirectional SH wave transducer based on a thickness poled piezoelectric ring. The ring was equally divided into twelve sectors and the electric field is circumferentially applied, resulting in the thickness-shear ($d_{15}$) mode. Both simulations and experimental testing results show that the proposed omnidirectional SH wave piezoelectric transducer (OSH-PT) can successfully excite and receive single-mode $SH_0$ wave at all directions with very small variations. The proposed OSH-PT is very promising in NDT and SHM due to its good performance, simple structure and low cost.

## 2. Structure and working principle of the proposed OSH-PT

To excite omnidirectional SH wave in a plate, an axisymmetric shearing force is required. For conventional thickness-shear ($d_{15}$) mode PZT transducers, the PZT is in-plane poled along the length direction ("3" direction) and the voltage is applied along the thickness direction ("1" direction ), as shown in figure 1(a). To realize an axisymmetric excitation force, the uniform circumferential polarization is required in a PZT ring. However, it is almost impossible to achieve uniform circumferential polarization in practical production as the PZT ring should be divided into several pieces and it is rather difficult to ensure the same polarization in all the PZT pieces. To avoid the extremely difficult manufacturing process, here we employed a new type of thickness-shear ($d_{15}$) mode to realize this particular shearing force. As shown in figure 1(b), the thickness-shear ($d_{15}$) mode can also be achieved when a PZT ceramics is poled along the thickness direction and the voltage is in-plane applied along the length direction. Compared with conventional $d_{15}$ mode PZT, the new $d_{15}$ mode PZT is much easier to fabricate since poling along the thickness direction is more favorable than poling along the length direction. The disadvantage of the new $d_{15}$ mode is that the voltage is applied along the length direction thus large electric field is difficult to apply, while this is not a problem in ultrasonic applications where only very small electric field is required.

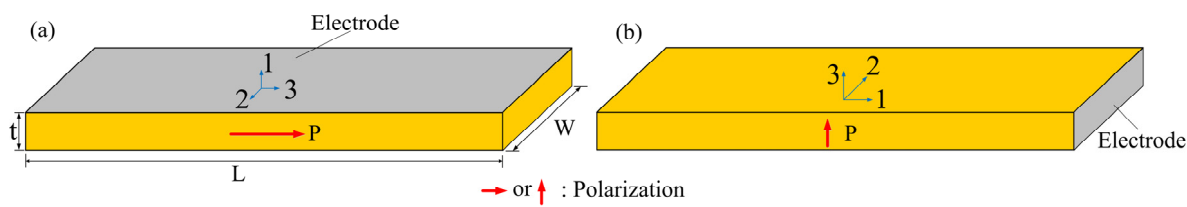

Figure 1. Two types of thickness-shear ($d_{15}$) mode in piezoelectric: (a) Conventional thickness-shear mode where poling is along length direction and field applied in the thickness direction. (b) The new thickness-shear mode where poling is along thickness direction and field applied along the length direction

The concept of this new $d_{15}$ mode is then applied on a PZT ring to realize an omnidirectional SH wave transducer. As shown in figure 2(a), a PZT-5H ring with the outer diameter of 21mm, inner diameter of 9mm and thickness of 2mm is firstly poled along the thickness direction. After poling, the ring is cut into twelve identical fan-shaped elements along its diameter. Then lateral electrodes were spread on each element to apply the electric field circumferentially. In order to keep the lateral faces bonded together being equi-potential for practical wiring, the poling direction of adjacent PZT elements is always opposite to each other for the ring transducer, as shown in figure 2(b). A single element with lateral electrodes is illustrated in figure 2(c), and figure 2 (d) is the photo of the actually-fabricated omnidirectional SH piezoelectric transducer (OSH-PT). Obviously for this OSH-PT, when a voltage is applied, circumferential electric field is input to the ring, resulting in a circumferential thickness-shear deformation. Therefore, when the OSH-PT is bonded on a large plate and an AC voltage is applied, alternative circumferential shearing force will be generated on the plate and omnidirectional SH wave is expected to be excited.

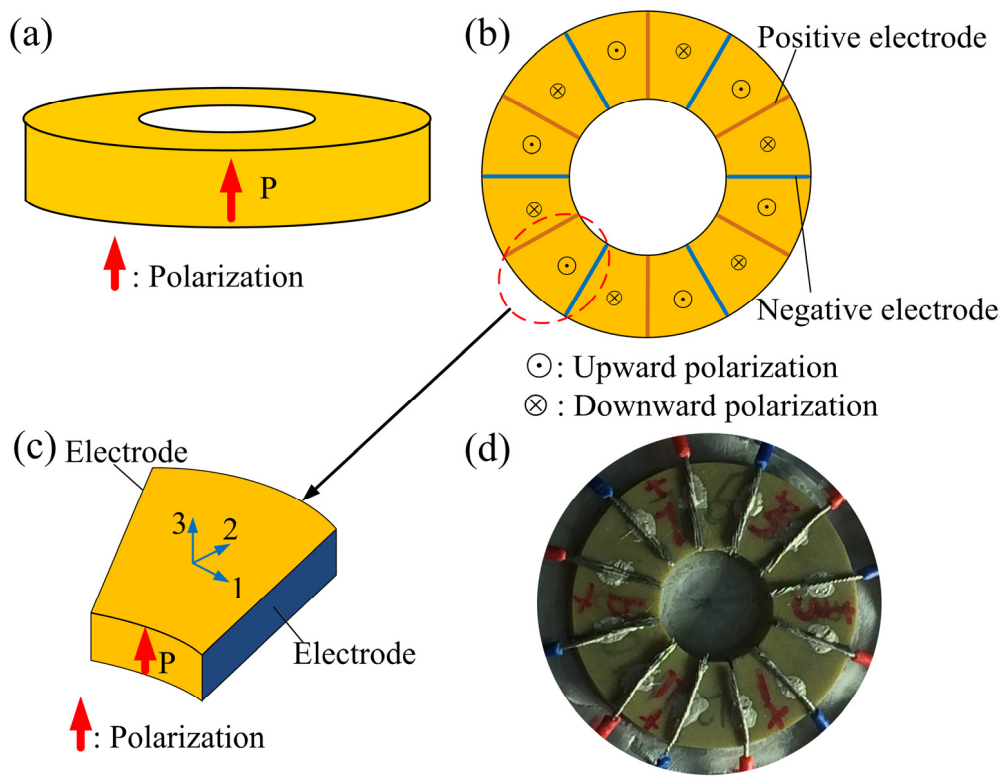

Figure 2. Schematic structure of the proposed omnidirectional SH wave piezoelectric transducer (OSH-PT): (a) a thickness poled PZT ring; (b) polarization and wiring of the 12 evenly divided elements of the PZT ring; (c) polarization and electrodes of a single PZT element; (d) photo of the actually-fabricated OSH-PT.

## 3. Finite element simulations

To examine the performance of the proposed OSH-PT in exciting and receiving SH waves, time-transient finite element (FEM) simulations were firstly conducted using ANSYS on an aluminum plate bonded with the proposed OSH-PT. The plate was modeled using SOLID 185 elements with the dimensions of 400 mm × 400 mm × 2 mm. The Young's modulus, Poisson ratio and density of it were 69 GPa, 0.33 and 2700kg·m$^{-3}$, respectively. The proposed OSH-PT was modeled using SOLID 5 elements and its dimensions are the same with that in Section 2. The material parameters of this PZT-5H transducer can be found elsewhere[25]. During simulation, the transducer was placed on the center of the aluminum plate. To ensure the accuracy of the computational results, the largest size of elements is less than 1/20 the shortest wavelength and the time step was less than 1/20 of the central frequency, as recommended in ANSYS User's Manual[26].

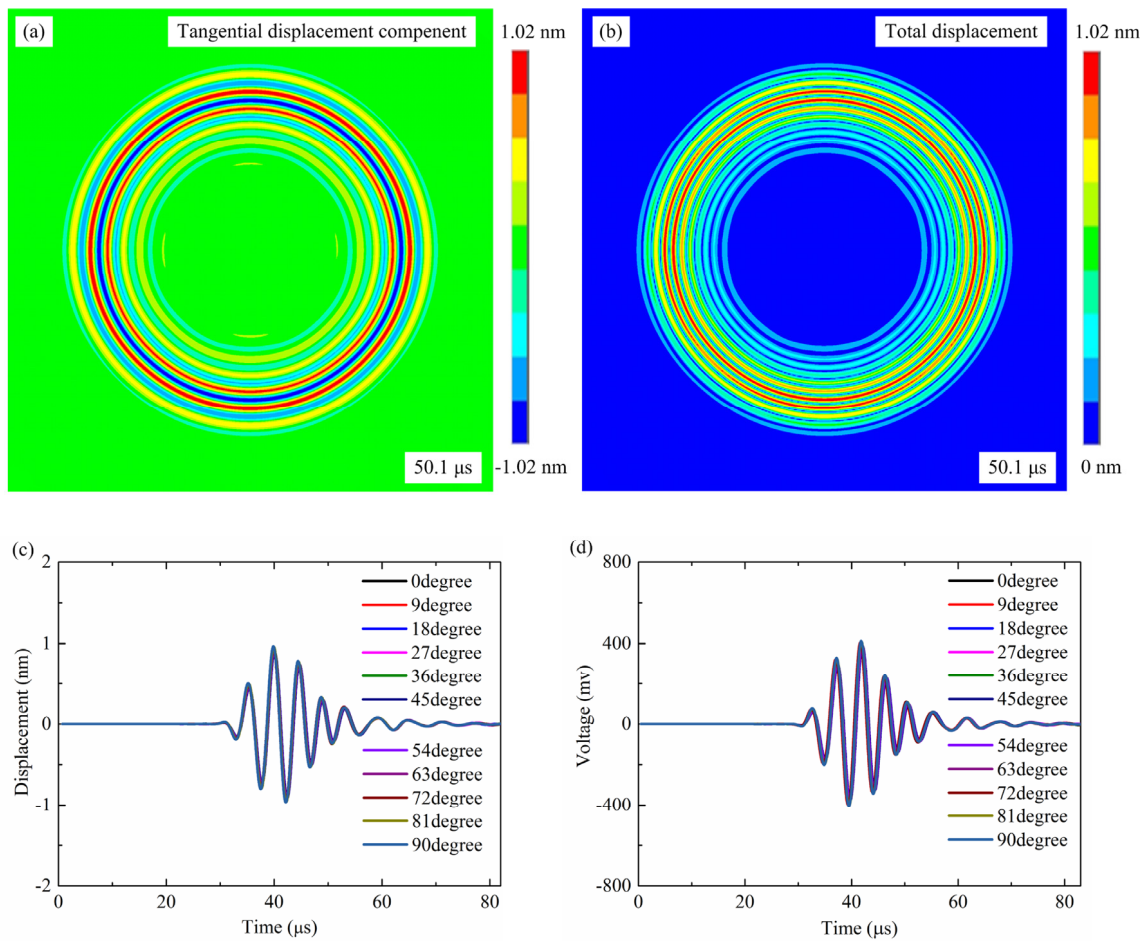

Figure 3. Finite element simulations of the proposed OSH-PT's performance on exciting and receiving SH waves in an aluminum plate. (a) Excited tangential displacement component at 50.1 μs, (b) Excited total displacement at 50.1 μs. (c) Time-domain tangential displacement excited by the proposed OSH-PT at different directions; (b) Time-domain voltage signals received by the proposed OSH-PT at different directions.

In the simulation, a five-cycle sinusoid tone-burst signal enclosed in a Hanning window was used to excite the proposed OSH-PT. The voltage of drive signal was set to be 20 V and the central frequency $f_c$ was fixed at 210 kHz. The snapshots of the tangential and total displacement fields under excitation of the OSH-PT at 50.1 μs were shown in figure 3(a) and figure 3(b), respectively. It can be seen in figure 3(a) that the tangential displacement component is perfectly axisymmetric, which representing a good omni-directivity of the $SH_0$ wave excited by the proposed OSH-PT. By referring to the amplitude of total displacement in figure 3(b), it can be concluded that the total displacement is totally dominated by the tangential displacement component, indicating that the amplitudes of the two lamb waves ($S_0$ and $A_0$) is negligible.

To further investigate the directivity of the proposed OSH-PT, time domain simulations were conducted using ANSYS where eleven monitoring points were positioned on the 1/4 circumference of a 100mm-radius circle with the OSH-PT at the center. As shown in figure 3(c), the tangential displacement signals of the excited $SH_0$ wave are nearly perfect five-cycle Hanning window-modulated sinusoid tone-burst. Furthermore, the curves for all the monitoring points were completely overlapped, indicating the perfect omni-directivity of the proposed OSH-PT in exciting $SH_0$ wave. When exploring the omni-directivity of proposed OSH-PT in receiving $SH_0$ wave, the tangential displacement modulated into the same window was applied at the eleven monitoring points respectively, and the proposed OSH-PT was used as a sensor. The voltage signal received by the OSH-PT was extracted in figure 3(d). Again, all the curves were completely overlapped, indicating the good omni-directivity of proposed OSH-PT in receiving SH wave. Note that all the results shown in figure 3 had theoretically confirmed the validity and omni-directivity of the proposed OSH-PT.

## 4. Experiment

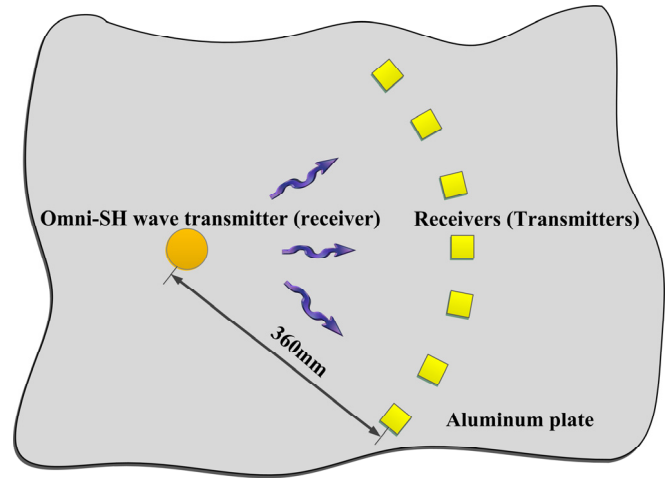

Figure 4. The schematic of experiments to check the wave mode and omni-directivity of the excited wave by the proposed OSH-PT.

Experiments were then designed to further investigate the performance of the proposed OSH-PT. An aluminum plate with the dimensions of 1000 mm × 1000 mm × 2 mm was used during testing. The proposed OSH-PT used here is the same as that used in the FEM simulations. To check the omnidirectional generation of $SH_0$ wave, the OSH-PT was used as a transmitter and the face-shear ($d_{36}$) type $0.72[Pb(Mg_{1/3}Nb_{2/3})O_3]$-$0.28[PbTiO_3]$ (PMN-PT) single crystal wafers (5 mm × 5 mm × 1 mm) were used as receivers. It should be noted that this type of PMN-PT wafers can excite and receive both $SH_0$ wave and Lamb waves since its piezoelectric coefficients $d_{31}$, $d_{33}$ and $d_{36}$ always coexist. Hence, the purity of $SH_0$ wave excited by the OSH-PT can be checked. The layout of experiments is illustrated in figure 4 where the OSH-PT is placed at the center of the plate and the $d_{36}$ type PMN-PT wafers were positioned on a circle at a distance of 360 mm from the OSH-PT with the angle spacing of 15°. Later, the $d_{36}$ type PMN-PT wafers were used as transmitters and the proposed OSH-PT was use as a receiver to examine its performance on receiving $SH_0$ waves. During testing, all the transmitters were driven by a five-cycle sinusoid tone-burst signal enclosed in a Hanning window using a function generator (3320A, Agilent, USA). A power amplifier (KH7602M) was used to amplify the drive signal and a digital oscilloscope (Agilent DSO-X 3024A) was used to collect the signals received by sensors.

4.1 Omnidirectional excitation of $SH_0$ wave by the proposed OSH-PT

Signals excited by the proposed OSH-PT and received by a $d_{36}$ type PMN-PT single crystal wafer were shown in figure 5(a). The drive voltage was set to be 20 V and its central frequency to be

210 kHz at first. As shown in figure 5(a), only $SH_0$ wave was successfully generated by the proposed OSH-PT. As expected, the waveform of the received $SH_0$ wave is almost identical to that of the drive signal with little distortion. Continuous wavelet transform (CWT) was then used to analyze both the drive and received signals. As shown in figure 5(b), the time interval between the drive signal and received signal was 118.5μs. Bearing in mind that the distance between the transmitter and the receiver is 360 mm, the measured group velocity of $SH_0$ wave in this aluminum plate is calculated to be 3051 m·s$^{-1}$, which is in good agreement with the theoretical value of 3099 m·s$^{-1}$.

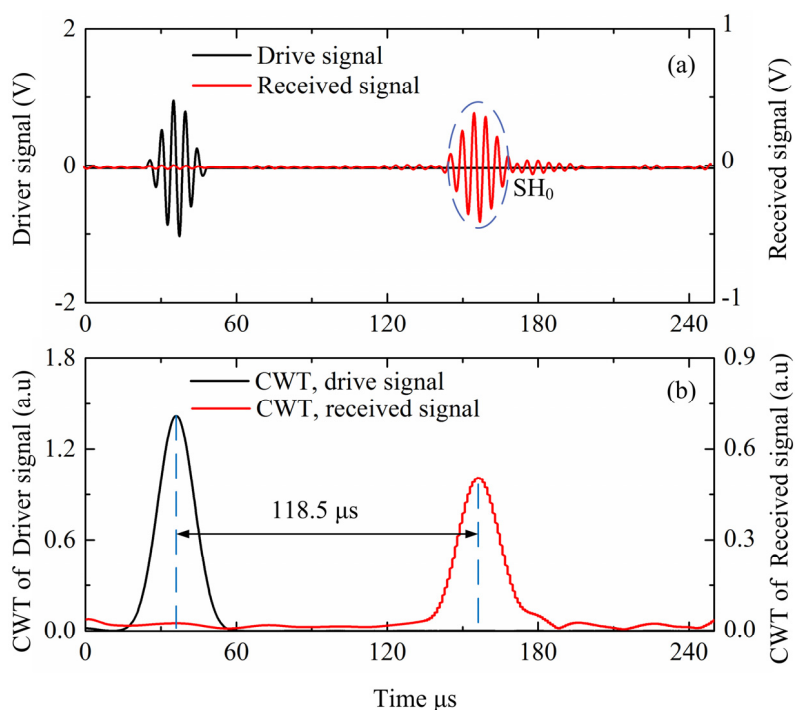

Figure 5. (a) Signals excited by the proposed OSH-PT at 210 kHz and received by a $d_{36}$ type PMN-PT single crystal wafer. (b) Continuous wavelet transform (CWT) of the drive signal and received signal.

Wave signals excited by the proposed OSH-PT at different frequencies under 20V and received by a $d_{36}$ type PMN-PT single crystal wafer were shown in figure 6. It can be seen that only $SH_0$ wave is received by the $d_{36}$ PMN-PT wafer in a wide frequency range from 110 kHz to 240 kHz. Since the $d_{36}$ PMN-PT wafers can detect both $SH_0$ wave and Lamb waves, these results indicates that the proposed OSH-PT can generate nearly pure $SH_0$ wave, as expected. From figure 6(a) to figure 6(e), it can be found that the amplitude of received $SH_0$ wave signals firstly increased monotonously with the increasing drive frequency, reaching its maxima at 210 kHz and then drop slightly with the increasing frequency. It should be mentioned that the amplitude of the received $SH_0$ wave can keep over 95% of the maximum value from 195 kHz to 225 kHz, which is attributed to the resonance effect of the transducer as well as the match between the wave length and the transducer's size.

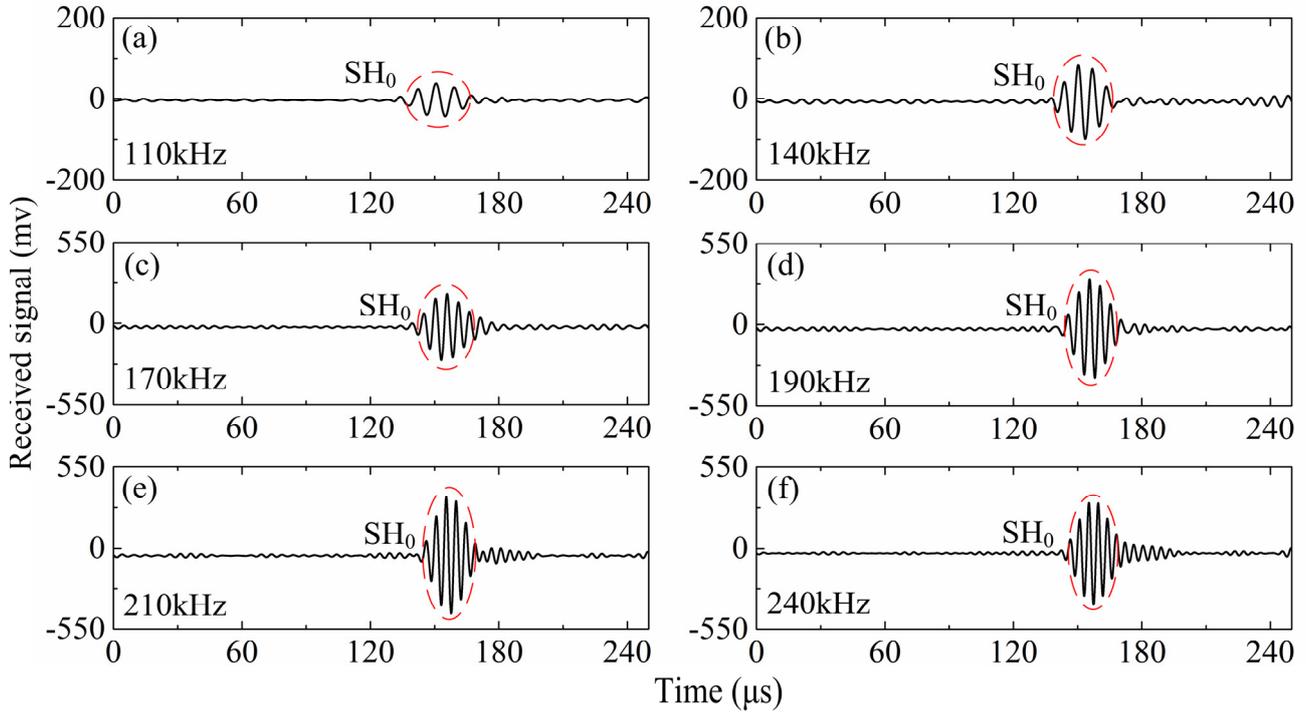

Figure 6. Wave signals excited by the proposed OSH-PT at different frequencies and received by a $d_{36}$ type PMN-PT single crystal wafer.

Figure 7 presented the omni-directivity of the proposed OSH-PT in generation of $SH_0$ wave. To validate the omni-directivity in a wide frequency range, the results in figure 7 were picked at 160 kHz, 195 kHz and 230 kHz, respectively. The peak-to-peak amplitudes of received $SH_0$ wave were normalized by average amplitude at the corresponding frequencies. Considering the axial-symmetry of the proposed OSH-PT, only the results from 0° to 90° were presented. As shown in figure 7, for the OSH-PT in generating $SH_0$ wave, the maximum amplitude deviation from the average amplitude was 3.8% at 160 kHz, 5% at 195 kHz, and 6.8% at 230 kHz, respectively, obviously smaller than that of about 15% for the previously developed OSH-PT based on artificial circumferentially poled face-shear (d24) PZT elements[24]. The variations in the PMN-PT receivers' properties and the non-uniform bond layers were thought to be responsible for the amplitude deviations at different directions. These non-uniformities could be avoided if measuring the wave field of the generated $SH_0$ wave by using a high-frequency 3-D laser scanning vibrometer[27]. Nevertheless, based on figure 6 and figure 7, it can be concluded that the proposed OSH-PT can generate omnidirectional $SH_0$ wave with nearly uniform sensitivity at all directions.

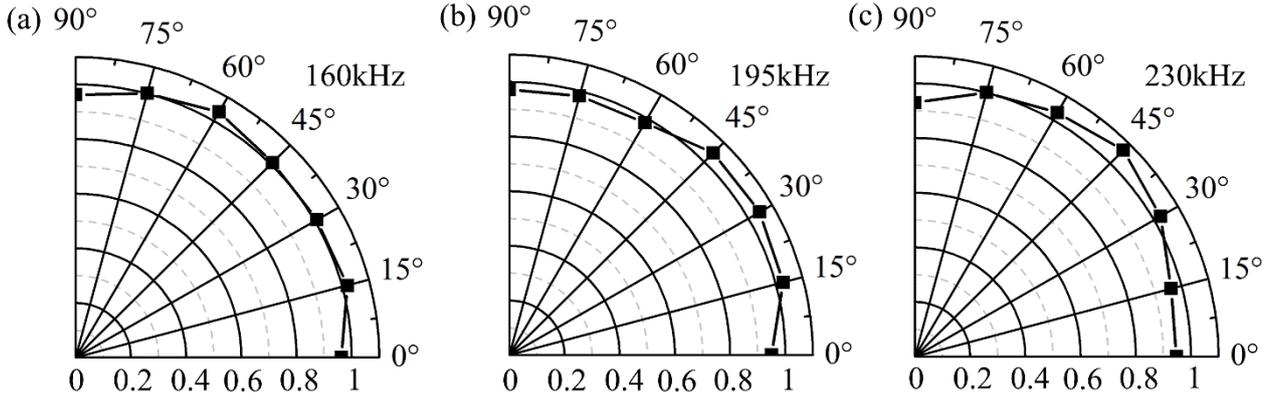

Figure 7. The omni-directivity of the proposed OSH-PT in generation of $SH_0$ wave at (a) 160 kHz, (b) 195 kHz and (c) 230 kHz. Wave signals were excited by the OSH-PT and received by $d_{36}$ type PMN-PT single crystal wafers. The amplitude is normalized by average amplitude at their respective frequencies.

4.2 Omnidirectional reception of $SH_0$ wave by the proposed OSH-PT

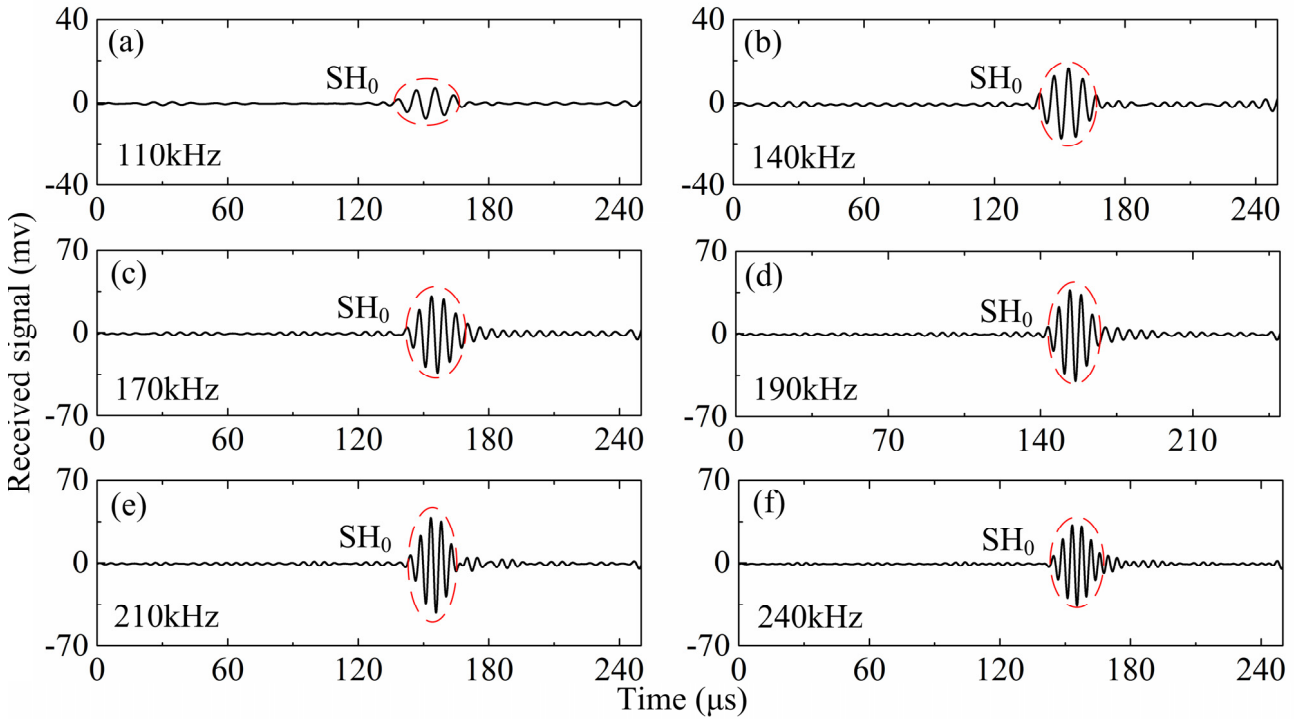

Figure 8. Wave signals excited by a $d_{36}$ type PMN-PT single crystal wafer at different frequencies and received by the proposed OSH-PT.

To examine the OSH-PT's performance in receiving SH wave, the $d_{36}$ type PMN-PT single crystal wafers were used as the transmitters and the proposed OSH-PT used as the sensor. The drive voltage was 20 V and the received signals at different frequencies were shown in figure 8. It can be seen that as expected, the $SH_0$ wave was successfully received by the proposed OSH-PT from 110 kHz to 240 kHz. It should be noted that the $d_{36}$ type PMN-PT single crystal wafer can generated both

SH$_0$ wave and Lamb waves. However, in figure 8, the OSH-PT only received the SH$_0$ wave and filtered the Lamb waves in a wide frequency range. This indicates that the proposed OSH-PT can realize wave filter inherently, which is very useful in assembling a phased array system.

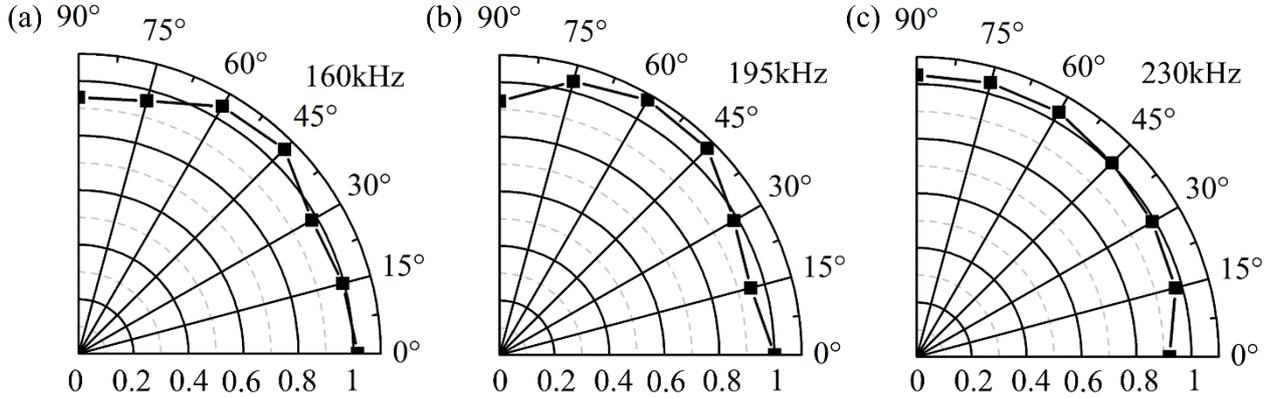

Figure 9. The omni-directivity of the proposed OSH-PT in reception of SH$_0$ wave at (a) 160 kHz, (b) 195 kHz and (c) 230 kHz. Wave signals were excited by d$_{36}$ type PMN-PT single crystal wafers and received by the OSH-PT. The amplitude is normalized by average amplitude at their respective frequencies.

Finally, the omni-directivity of the proposed OSH-PT in reception of SH$_0$ wave was investigated. Similarly, the received signals were also picked at 160 kHz, 195 kHz and 230 kHz to check the omni-directivity in a wide frequency range and the amplitudes were normalized by their average amplitude at respective frequencies, as shown in figure 9. Also as expected, the proposed OSH-PT shows nearly uniform omni-directivity in reception of SH$_0$ wave. The maximum deviation from the average amplitude was 5.8% at 160 kHz, 7.3% at 195 kHz, and 7.6% at 230 kHz, respectively, which are also considerably smaller than that of ~15% for the previously developed OSH-PT based on face-shear (d$_{24}$) PZT elements[24]. The deviation here can also be attributed to the variations in the different PMN-PT wafers, which can only be diminished by using strictly identical transmitters at all directions. Anyway, the results in figure 8 and figure 9 indicated that the proposed OSH-PT has a good performance in receiving SH$_0$ wave with nearly uniform sensitivities at all directions.

## 4. Conclusion

In summary, an omnidirectional SH wave piezoelectric transducer (OSH-PT) is proposed based on a thickness poled, thickness-shear piezoelectric ring. The ring was cut to twelve identical elements and circumferential electric field was applied, resulting in the thickness-shear (d$_{15}$) mode. Since all the elements composing the proposed OSH-PT came from the same thickness poled PZT ring, the variations in different elements were minimized. The performance of the proposed OSH-PT was validated by both FEM simulations and experimental testing. All the results indicated that the proposed OSH-PT can generate and receive single mode of $SH_0$ wave in a wide frequency range. Furthermore, the omni-directivity of the proposed OSH-PT is also very good whether it acts as a SH wave transmitter or receiver. Considering its compact size, simple structure and low cost, the proposed OSH-PT is very promising in the field of NDT and SHM. In our ongoing work, we focus on developing a phased array system consisting of many OSH-PTs for quick scan of large plate-like structures.